\documentclass[pdflatex,sn-mathphys-num]{sn-jnl}

\usepackage{graphicx}%
\usepackage{multirow}%
\usepackage{amsmath,amssymb,amsfonts}%
\usepackage{amsthm}%
\usepackage{mathrsfs}%
\usepackage[title]{appendix}%
\usepackage{xcolor}%
\usepackage{textcomp}%
\usepackage{manyfoot}%
\usepackage{booktabs}%
\usepackage{algorithm}%
\usepackage{algorithmicx}%
\usepackage{algpseudocode}%
\usepackage{listings}%

\theoremstyle{thmstyleone}%
\theoremstyle{thmstyletwo}%

\theoremstyle{thmstylethree}%

\begin{document}

\title[Article Title]{\textbf{A Green Solution for Breast Region Segmentation Using Deep Active Learning}}


\author*[ 1,2,3]{\fnm{Sam} \sur{Narimani}}\email{sam.narimania@gmail.com }

\author[ 3,4]{\fnm{Solveig} \sur{Roth Hoff}}\email{solveig.roth.hoff@ntnu.no}


\author[ 5,6]{\fnm{Kathinka} \sur{Dæhli Kurz}}\email{kathinka.dehli.kurz@sus.no}

\author[ 7, 8]{\fnm{Kjell-Inge} \sur{Gjesdal}}\email{kjell.inge.gjesdal@nordiccad.com}

\author[ 9,10]{\fnm{Jürgen} \sur{Geisler}}\email{jurgen.geisler@medisin.uio.no}

\author[ 1,2,3]{\fnm{Endre} \sur{Grøvik}}\email{endre.grovik@gmail.com}

\affil[1]{\small{\orgdiv{ Department of Physics}, \orgname{Norwegian University of Science and Technology}, \orgaddress{ \city{Trondheim},  \country{Norway}}}}

\affil[2]{\small{\orgdiv{ Research and Development Department}, \orgname{More og Romsdal Hospital Trust}, \orgaddress{ \city{Aalesund},  \country{Norway}}}}

\affil[3]{\small{\orgdiv{ Department of Radiology}, \orgname{More og Romsdal Hospital Trust}, \orgaddress{ \city{Aalesund},  \country{Norway}}}}

\affil[4]{\small{\orgdiv{ Department of Health Sciences}, \orgname{Norwegian University of Science and Technology}, \orgaddress{ \city{Aalesund},  \country{Norway}}}}

\affil[5]  {\small {\orgdiv{ Stavanger Medical Imaging Group, Radiology Department}, \orgname{Stavanger University Hospital}, \orgaddress{ \city{Stavanger},  \country{Norway}}}}

\affil[6]  {\small {\orgdiv{ Department of Electrical Engineering and Computer Science}, \orgname{The University of Stavanger}, \orgaddress{ \city{Stavanger},  \country{Norway}}}}

\affil[7] {\small  \orgdiv{ Department of Diagnostic Imaging}, \orgname{Akershus University Hospital}, \orgaddress{ \city{Lorenskog},  \country{Norway}}}
\affil[8] {\small \orgname{ NordicCAD AS}, \orgaddress{ \city{Aalesund},  \country{Norway}}}
\affil[9] {\small  \orgdiv{ Institute of Clinical Medicine}, \orgname{Faculty of Medicine}, \orgaddress{ \city{University of Oslo},  \country{Norway}}}

\affil[10] {\small  \orgdiv{ Akershus University Hospital}, \orgname{Department of Oncology}, \orgaddress{ \city{Lorenskog},  \country{Norway}}}




\abstract{\textbf{Purpose:} Annotation of medical breast images is an essential step toward better diagnostic but a time consuming task. This research aims to focus on different selecting sample strategies within deep active learning on Breast Region Segmentation (BRS) to lessen computational cost of training and effective use of resources.

\textbf{Methods:}  The Stavanger breast MRI dataset containing 59 patients was used in this study, with FCN-ResNet50 adopted as a sustainable deep learning (DL) model. A novel sample selection approach based on Breast Anatomy Geometry (BAG) analysis was introduced to group data with similar informative features for DL. Patient positioning and Breast Size were considered the key selection criteria in this process. Four selection strategies including Random Selection, Nearest Point, Breast Size, and a hybrid of all three strategies were evaluated using an active learning framework. Four training data proportions of 10\%, 20\%, 30\%, and 40\% were used for model training, with the remaining data reserved for testing. Model performance was assessed using Dice score, Intersection over Union, precision, and recall, along with 5-fold cross-validation to enhance generalizability.
 
\textbf{Results:} Increasing the training data proportion from 10\% to 40\% improved segmentation performance for nearly all strategies, except for Random Selection. The Nearest Point strategy consistently achieved the lowest carbon footprint at 30\% and 40\% data proportions. Overall, combining the Nearest Point strategy with 30\% of the training data provided the best balance between segmentation performance, efficiency, and environmental sustainability.

}

\keywords{Deep Active Learning, Breast Region Segmentation, Human-center analysis}

\maketitle

\section{Introduction}\label{sec1}

Segmentation of medical imaging is  a crucial step in diagnostics and treatment process \cite{xu2024advances}. Whole breast segmentation, for instance, can lead to quantitative analysis of breast anatomy and facilitating for better breast tumor segmentation \cite{acciavatti2023beyond,lew2024publicly,xu2018breast}.
However accurate annotation of the medical images like Magnetic Resonance (MR) breast images needs expertise and special skills \cite{budd2021survey}. Precisely drawing around the Region of Interest (ROI) on all slices is an essential step for supervised learning, but is a time consuming task. \cite{li2023hal}. On the other hand, carbon emission due to training of big data is one of the reasons for global warming and may represent a threat for the future of the earth \cite{zhong2024impact}.

Breast region segmentation (BRS) is a desired step prior to breast lesion segmentation to lessen computational cost and accurately segment the lesion in MR breast images \cite{narimani2025sustainable}. Deep learning (DL) methods have demonstrated promising results in segmentation of MR breast regions \cite{pandey2018automatic, narimani2025comparative} and density \cite{zhang2021development} in the last decade. Advancement in eco-friendly DL models has paved the way for more sustainable methods.  In a recent study, exploring seven DL models ability to segment the defined breast region with a novel method for breast boundary, the model FCNResNet50 was found to be the most relevant, eco-friendly and sustainable model. \cite{narimani2025comparative}. In addition to sustainable DL models, other attractive methods related to uncertain and informative data can be hired to decrease annotation burden in DL training process \cite{fu2023attention}.

Active learning is one of the powerful approaches in Artificial Intelligence (AI) to optimize resources by introducing strategies to select the most uncertain and informative data for labeling \cite{munro2021human, haug2021applying}. Common segmentation approaches in medical imaging often rely on supervised learning techniques requiring large labeled datasets for training \cite{ajiboye2015evaluating}. In addition, the manual segmentation of medical images with accurate pixel-level labels is challenging and can introduce variability and subjectivity in the segmentation process \cite{rajchl2017employing}. Furthermore, the scarcity of annotated data in breast imaging datasets pose significant challenges in developing robust segmentation models for breast region delineation \cite{weese2016four}. Active learning, an iterative learning paradigm, has emerged as a promising solution to address these challenges in whole breast segmentation. By strategically selecting and prioritizing the most valuable and uncertain data points for annotation, active learning enables the efficient utilization of limited annotated data and resources while optimizing model performance.

The aim of this study is to find Sample Selection Strategy (SSS) to primary lessen amount of data for DL model training. In fact by mitigation of training dataset size, we not only reach to a more environmental friendly approach but also a sustainable approach for future research in segmentation of MR breast images.

\section{Materials and Methods}\label{sec2}

\subsection{Data and preprocessing}\label{subsec2}

The breast MRI dataset used in this study comprises images from 59 patients acquired in 2008 at Stavanger University Hospital. Image acquisition was performed using a 1.5 T Philips MRI machine equipped with a dedicated breast coil. Each examination consisted of six scans, one pre-contrast and five post-contrast series. A detailed overview of the imaging protocol, acquisition parameters, and dataset characteristics is available in our previous study \cite{narimani2025comparative}.
All scans were obtained with patients positioned Head-First Prone (HFP) and oriented according to the Right–Anterior–Superior (RAS) coordinate system, which represents the standard orientation in Scandinavian imaging practice. The original DICOM (Digital Imaging and Communications in Medicine) images were converted to NIfTI (Neuroimaging Informatics Technology Initiative) format to facilitate preprocessing and subsequent model training. Standard preprocessing procedures such as oversampling and image size normalization were applied to ensure spatial uniformity and consistency across all patients. During this stage, one patient’s data were excluded due to a mismatch between the pre-contrast and first post-contrast scans.

\subsection{Active learning}\label{subsec2}

\begin{figure}[h!]
\centering
\includegraphics[width=1.05\textwidth, trim=0 0 0 0 , clip]{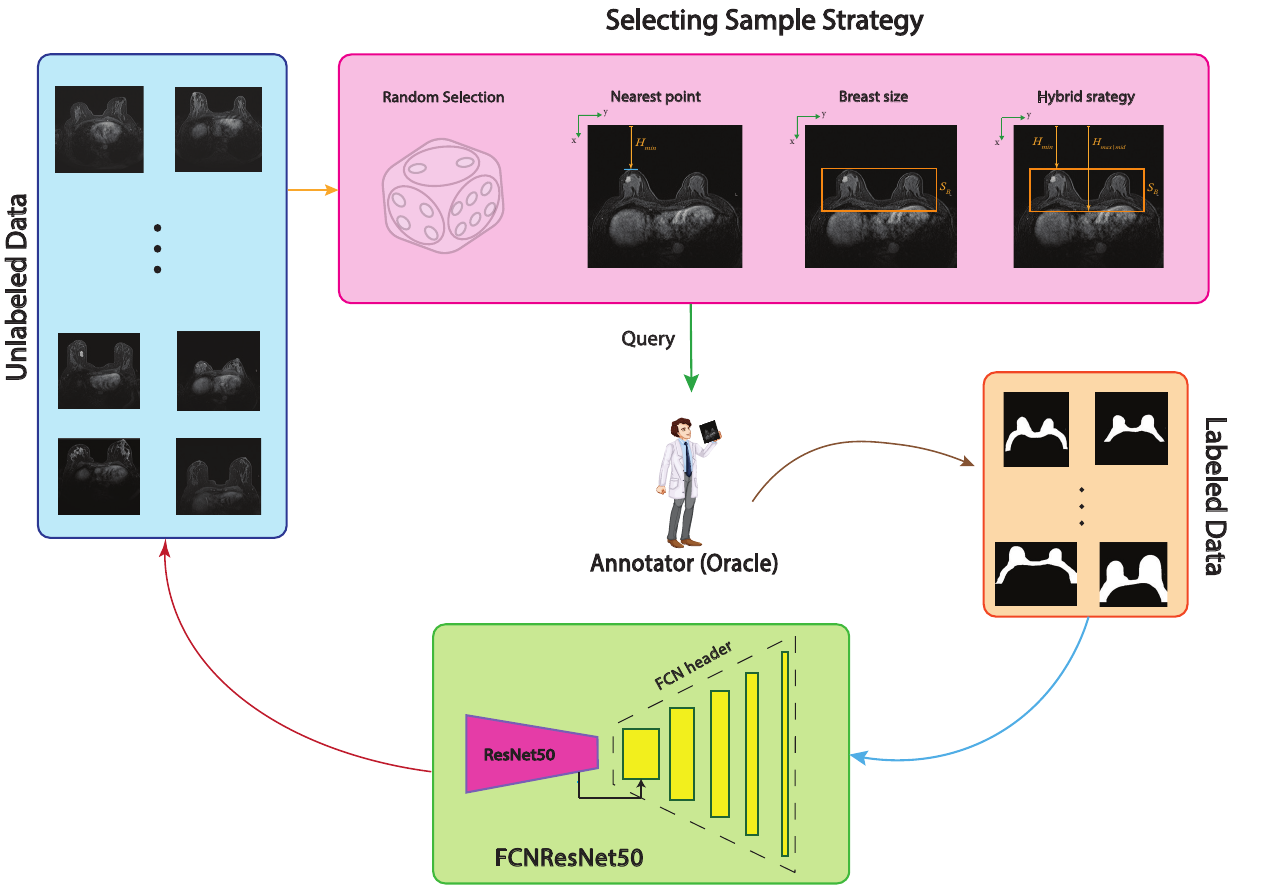}
\caption{\centering Schematic illustration of the proposed active learning cycle  }\label{Paperdesign}
\end{figure}

Active learning is a process for determining which queries, samples, or data points should be annotated within a human-in-the-loop framework \cite{mosqueira2023human}. Various strategies are employed to evaluate the efficiency of an active learning pipeline, and the choice of strategy often plays a crucial role in the overall performance of the system. As shown this process in the  Figure \ref{Paperdesign}, a subset of data samples is selected based on a SSS and then provided to an AI model for training. Before training, the selected data are labeled by skilled human annotators with domain expertise. The labeled data are subsequently used to train and fine-tune the model to achieve the best fit. Once trained, the model predicts labels for the remaining unlabeled data pool to assess the effectiveness of the active learning pipeline and its underlying strategy. If the predefined performance criteria are not met, the model’s predictions are reviewed, and the uncertain samples are either reintroduced into the unlabeled data pool or manually annotated by Oracles or human experts. This iterative cycle continues until all unlabeled data are annotated either automatically by the model or manually by annotators. The following sections provide a detailed explanation of the individual components of the active learning process.

\subsubsection{Sample Selection Strategy}\label{subsubsec2}
Identifying the most informative and uncertain samples is the core principle of the active learning framework, as it enables the model to achieve high performance with the minimum amount of annotated data \cite{lewis1994sequential, settles2012active}. This process is implemented through SSS, which are broadly categorized into three main types namely random sampling, uncertainty, and diversity. Random sampling is the most common and widely used strategy in artificial intelligence applications, particularly for splitting datasets into training and testing sets \cite{liu2021influence-random}. However, its main limitation is that it does not guarantee the selection of informative or representative samples, which may result in the exclusion of valuable data and consequently degrade model performance. In contrast, uncertainty-based sampling selects instances for which the model’s predictions are most uncertain, typically those that lie near the decision boundaries between classes \cite{nath2020diminishing-uncertainty}. These samples are difficult to classify and therefore provide the most informative feedback for improving the model. Finally, diversity-based methods aim to identify rare or underrepresented samples in the dataset, or samples that effectively represent different regions of the data distribution \cite{jin2022one-diversity}. These approaches ensure broader coverage of the input space and reduce redundancy in selected data. Overall, active learning seeks to acquire new, previously unseen data that contain highly informative features, thereby maximizing learning efficiency while minimizing annotation costs.

\subsubsection*{Breast Anatomy Geometry Analysis}

\begin{figure}[h!]
\centering
\includegraphics[width=1.15\textwidth, trim=0 30 0 0 , clip]{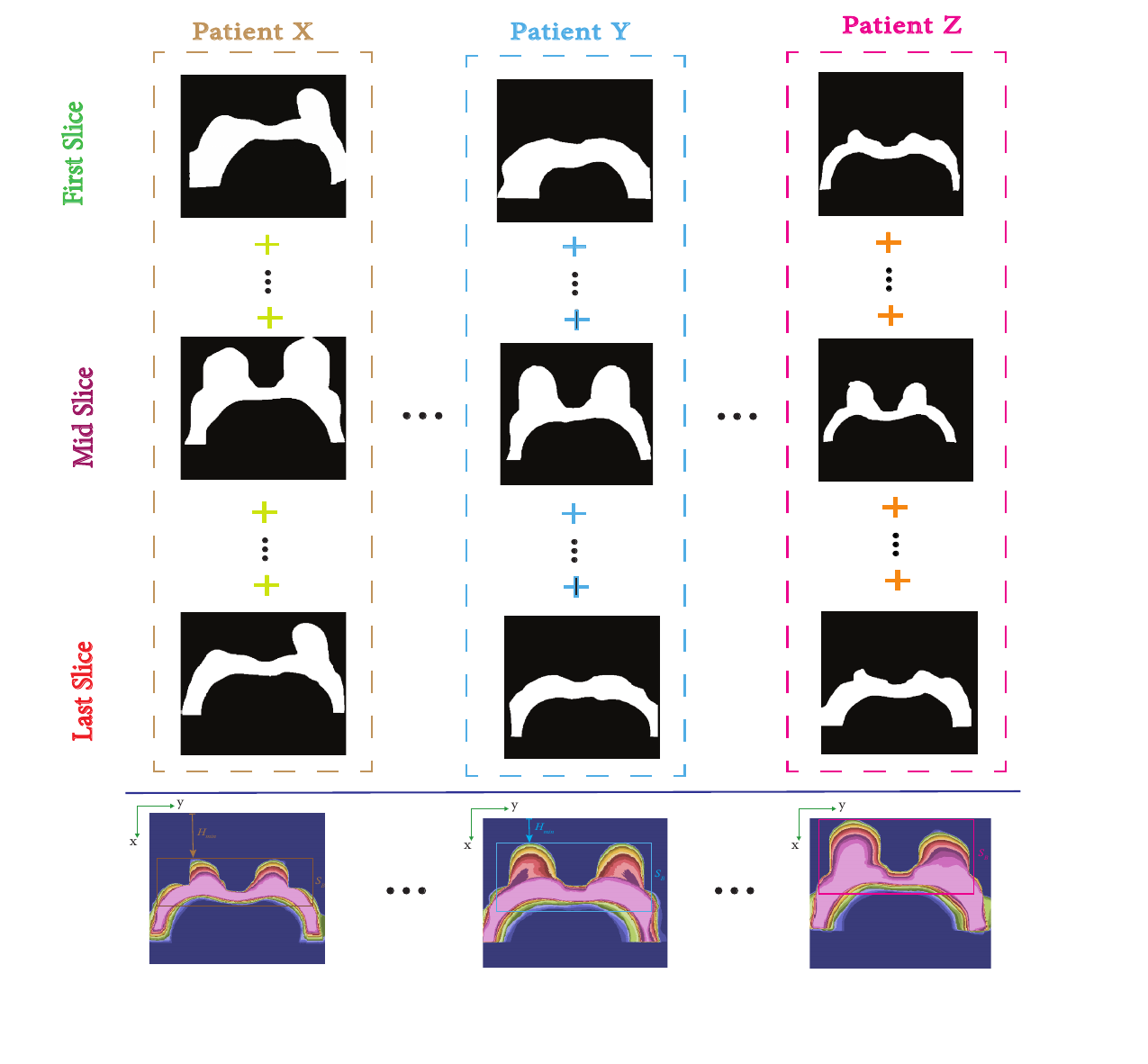}
\caption{\centering Illustration of different breast anatomy geometries including Nearest Point and Breast Size by using breast region overlay map }\label{BAG}
\end{figure}

As demonstrated in our previous studies \cite{narimani2025comparative}, accurate delineation of breast boundaries plays a critical role in improving the performance of breast lesion segmentation. The extraction of breast boundaries can be exploited in two primary ways. First, defining the breast region enables the analysis of breast region distribution, revealing where breast tissue is most frequently located within the image. Second, boundary information facilitates the estimation of Breast Size in a straightforward and reliable manner. Analysis of breast MR images further indicates that breast positioning, which is closely related to patient positioning, varies substantially according to individual anatomical characteristics. For instance, the nipple location may appear near the upper edge of the image in some patients, while in others it is approximately one-third below the vertical (y-axis) coordinate. This anatomical variability suggests that clustering breast MR images based on boundary-derived features can lead to a more meaningful categorization of the dataset. From each cluster, a representative sample can then be selected for training process based on the first boundary contact identified through data-driven analysis. An alternative complementary strategy involves the analysis of Breast Size in MR images, where accurate breast boundary extraction allows precise Breast Size estimation. Figure \ref{BAG} illustrates breast slices from multiple patients, highlighting the substantial variation in breast positioning across different MR image volumes. Furthermore, by constructing a breast region overlay map obtained by summing the segmented breast regions across all slices of a given patient, the overall spatial distribution of breast tissue can be visualized more effectively. As shown in the figure, both the distance of the nearest breast boundary point from the image coordinate origin and the Breast Size vary considerably among patients. Consequently, these two factors, breast position and Breast Size, form the fundamental components of our Breast Anatomy Geometry (BAG) analysis framework.

\subsection{Deep learning Architecture}\label{subsec2}
\begin{figure}[h!]
\centering
\includegraphics[width=1.05\textwidth, trim=0 70 0 0 , clip]{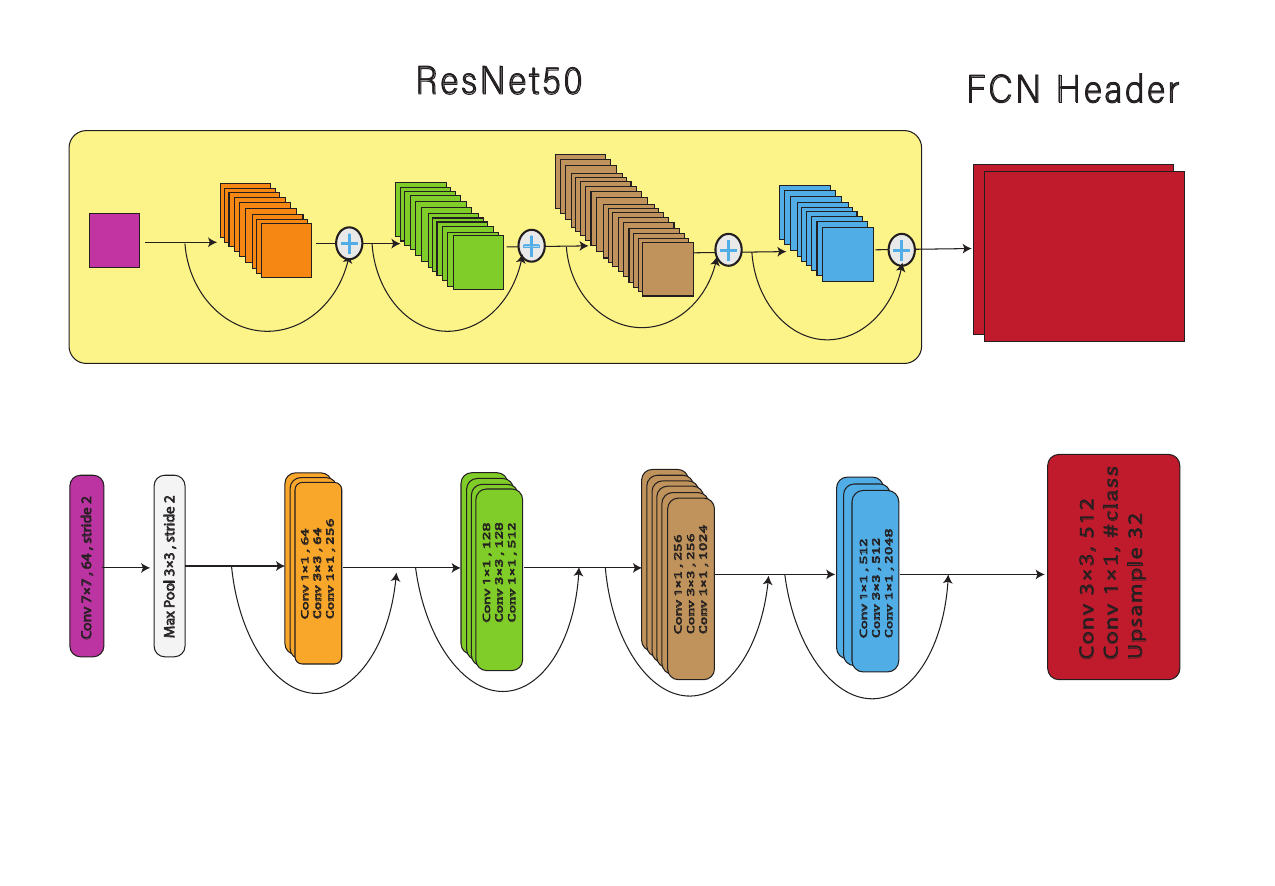}
\caption{\centering Block diagram of the FCN-ResNet50 architecture showing specifications of each layer }\label{fcnresnet50}
\end{figure}

As suggested for sustainable DL-based BRS \cite{narimani2025comparative}, FCN-ResNet50 demonstrated competitive segmentation performance while achieving the lowest carbon footprint among the evaluated models. Consequently, this environmentally sustainable segmentation network is adopted as the core model within our active DL framework.
FCN-ResNet50 consists of an encoder–decoder architecture composed of a ResNet50 backbone as the encoder and an FCN-based decoder head. ResNet has consistently shown strong performance in feature extraction across a wide range of segmentation networks \cite{he2016deep-resnet}. Meanwhile, the FCN head serves as an efficient upsampling decoder, particularly effective for segmenting large anatomical structures \cite{long2015fully-fcn}. The combination of these two components makes this model well suited for whole-breast segmentation tasks.
Figure \ref{fcnresnet50} depicts the detailed architecture of the proposed DL model, including layer-wise specifications.

\subsection{Evaluation}\label{subsec2}

Model performance was evaluated through two distinct processes namely training and testing. During the training phase, the Dice loss function was employed to support efficient optimization and faster convergence of the DL model. In addition, five-fold cross-validation was implemented to improve the reliability and generalizability of the results, given the relatively small size of the dataset. In contrast, model performance on unseen data was assessed using several evaluation metrics, including Dice score, Intersection over Union (IoU), Precision, Recall and Hausdorff Distance (HD). Dice loss, Precision , Recall, IoU  and HD are formulated as equations \ref{eq-DiceLoss} to \ref{eq-HD}, respectively:

\begin{equation}
\mathcal{L}_{Dice} = 1 - \frac{2 \sum_{i} p_i g_i + \epsilon}{\sum_{i} p_i + \sum_{i} g_i + \epsilon}
\label{eq-DiceLoss}
\end{equation}

\begin{equation}
\text{Precision} = \frac{\sum_i (p_i \cdot g_i + \epsilon)}{\sum_i p_i + \epsilon}
\label{eq-Precision}
\end{equation}

\begin{equation}
\text{Recall} = \frac{\sum_i (p_i \cdot g_i + \epsilon)}{\sum_i g_i + \epsilon}
\label{eq-Recall}
\end{equation}

\begin{equation}
\text{IoU} = \frac{\sum_i (p_i \cdot g_i + \epsilon)}{\sum_i (p_i + g_i - p_i \cdot g_i + \epsilon )}
\label{eq-IoU}
\end{equation}

\begin{equation}
d_H(A, B) = \max \left\{
\sup_{a \in A} \inf_{b \in B} \|a - b\|,
\sup_{b \in B} \inf_{a \in A} \|b - a\|
\right\}
\label{eq-HD}
\end{equation}

\vspace{.5cm}
where $p_i \in \{0,1\}$ and $g_i \in \{0,1\}$ represent the predicted label  and corresponding ground-truth label for pixel $i$, respectively. A small constant $\epsilon$ is added to both the numerator and denominator to prevent division by zero and to ensure numerical stability during training and testing. In addition, A and B denote the predicted and ground-truth surfaces, respectively, while a and b represent arbitrary points belonging to A and B.

On the other hand, the carbon footprint is a critical factor in AI applications that must be considered \cite{strubell2020energy}. The average carbon footprint for producing 1 kWh of energy is reported to be 475 grCO2 \cite{iea2019}. Consequently, the carbon footprint (CFP) for each fold can be calculated using relation \ref{eq-sigma}:

\begin{equation}
\text{CFP} = \epsilon_f \cdot \sum_{i=1}^{N} P_i \frac{\Delta t}{3600}
\label{eq-sigma}
\end{equation}

\noindent
where $\epsilon_f = 0.475~\mathrm{kg/kWh}$ is the emission factor \cite{iea2019}, $P_i$ denotes the power in watts at time step $i$, and $\Delta t$ is the duration of each time step in seconds. The division by $3600$ converts seconds to hours. Eventually, \(\text{CFP}\) represents the carbon footprint in kilograms of CO2 for each fold during each training session.

\section{Results}\label{sec3}
\subsection{Experiments}\label{sec3}
Four different approaches, illustrated in Figure \ref{Paperdesign}, were employed to provide input data to the DL model. In addition to these strategies, the proportion of data used for training was varied from 10{\%} to 40{\%}, with a 10{\%} decrement at each step to assess the model's performance. Consequently, a total of 16 distinct models were trained to comprehensively evaluate model efficiency.
The model hyperparameters and input specifications are presented in Table \ref{tab:training_config}. As shown in the table, the input consists of pre-contrast and first post-contrast images, fed to the model as 2D slices during the training process. The corresponding mask file for each input, manually annotated, serves as the ground truth output in our active learning cycle. The FCN-ResNet50 architecture used in this study was modified to accommodate our specific task, as the input comprises only two channels (pre-contrast and first post-contrast) and a single output class, rather than the original configuration with 21 classes. Therefore, minor adjustments were applied to the base architecture to tailor it to the requirements of this study.
Eventually a five-fold cross-validation strategy was adopted, not only to ensure robust performance evaluation given the relatively small dataset but also to reduce computational cost by training on smaller folds.

\begin{table}[ht]
\centering
\caption{\centering Summary of training configuration and parameters}
\label{tab:training_config}
\begin{tabular}{ll}
\hline
\textbf{Parameter / Aspect} & \textbf{Description} \\
\hline
Input Data & Pre- and first post-contrast images \\
Data Format & NIfTI images  \\
Training Approach & Slice-by-slice training \\
Model Type & FCNResNet50 \\
Validation Strategy & 5-fold cross-validation. \\
Loss Function & Dice loss function. \\
Optimizer & RAdam optimizer. \\
Initial Learning Rate & 0.001 \\
Learning Rate Scheduler & \textit{ReduceLROnPlateau} \\
Batch Size & 8 \\
Data Shuffling & Utilized only in the model training stage \\
\hline
\end{tabular}
\end{table}

\subsection{Breast Anatomy Geometry}

\begin{figure}[h!]
\centering
\includegraphics[width=1\textwidth, trim=0 0 0 0 , clip]{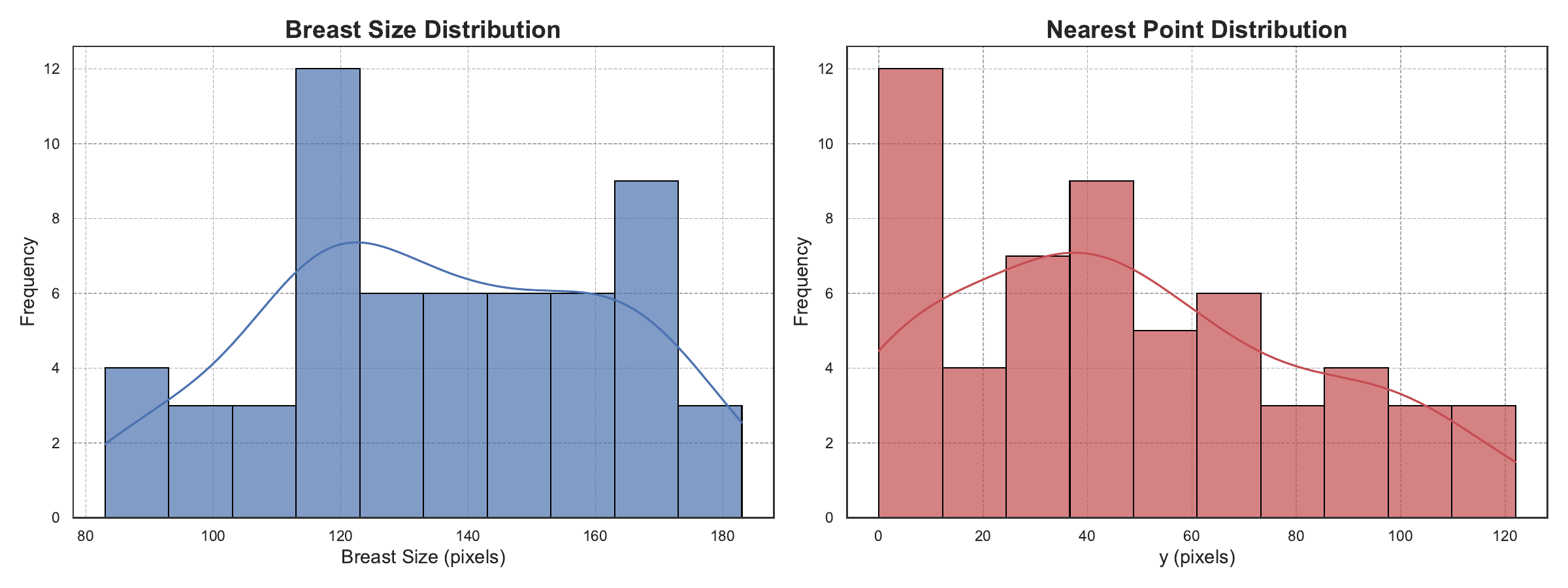}
\caption{\centering Breast Size and Nearest Point histogram along all patients}\label{histogram_analysis}
\end{figure}

The distribution of the Nearest Point along the y-coordinate and Breast Size exhibits variability among patients, reflecting differences in individual physical characteristics. As shown in Figure \ref{histogram_analysis}, Breast Sizes are predominantly clustered around 120 pixels. However, a minority of cases display notably smaller or larger Breast Sizes, which may influence the generalizability of the dataset. Furthermore, twelve breast images are positioned proximate to the top of the coordinate range, resulting in the ROI occupying the upper portion of these images. This pattern suggests the existence of cases with greater deviations from the y-coordinate, potentially corresponding to smaller breast volumes or individuals with smaller body frames.

\subsection{Strategy performance }\label{sec3}
\begin{figure}[h!]
\centering
\includegraphics[width=1\textwidth, trim=0 0 0 0 , clip]{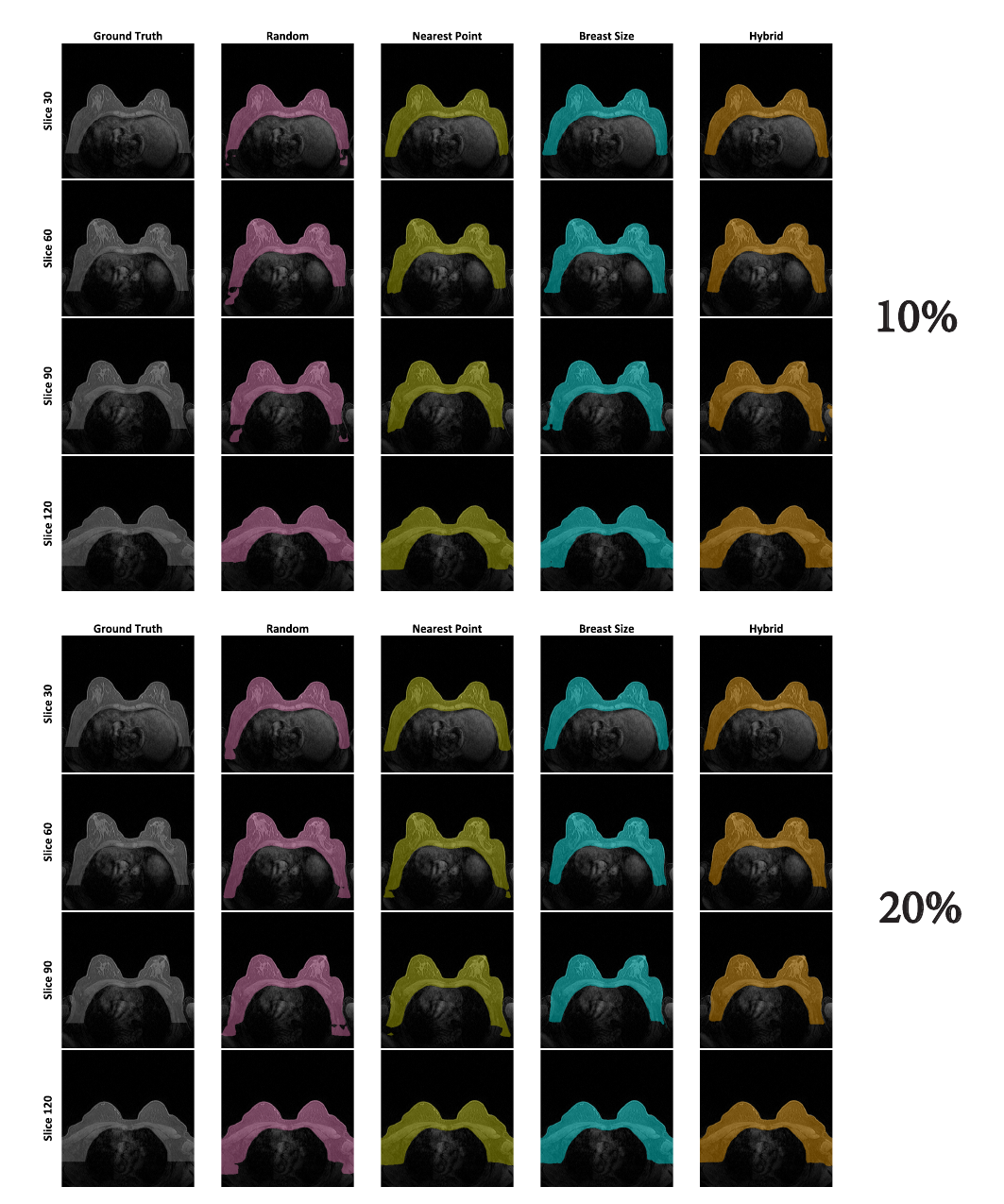}
\caption{\centering Segmentation results across all strategies for different slices for test dataset by using model trained with 10 and 20 percent of whole dataset}\label{10-20_seg_result}
\end{figure}

\begin{figure}[h!]
\centering
\includegraphics[width=1\textwidth, trim=0 0 0 0 , clip]{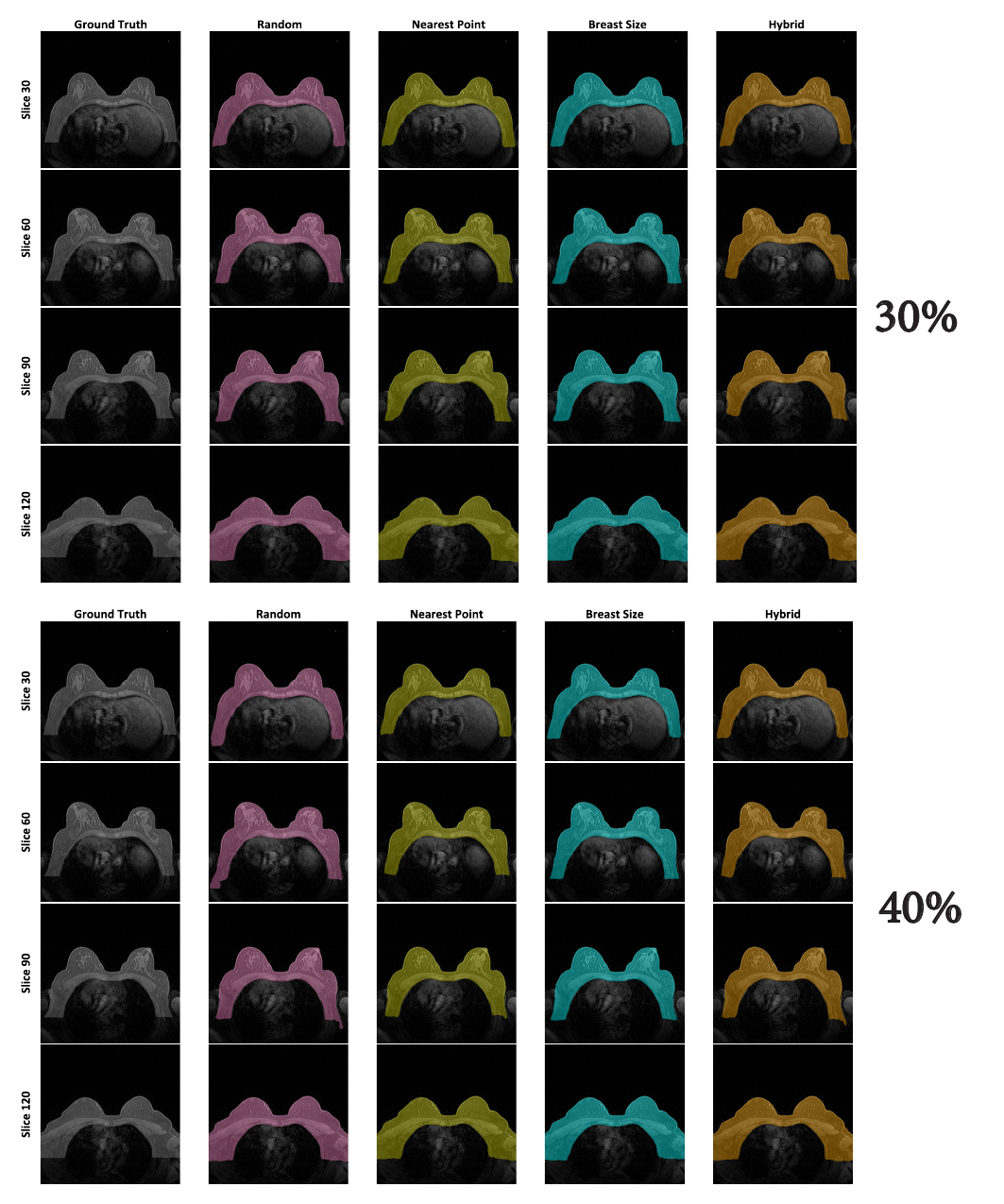}
\caption{\centering Segmentation results across all strategies for different slices for test dataset by using model trained with 30 and 40 percent of whole dataset}\label{30-40_seg_result}
\end{figure}

Different segmentation results for the four strategies are shown in Figures \ref{10-20_seg_result} and \ref{30-40_seg_result} using the test data, to which none of the models had been exposed during training. As illustrated in Figure \ref{10-20_seg_result}, the segmentation results with 10\% and 20\% of the training data show satisfactory internal segmentation, which is crucial for the subsequent lesion segmentation step. For boundary accuracy and overall performance, the Random Selection strategy produces poorer results compared to the other strategies. However, as the proportion of training data increases, the model’s predictions improve even for the random strategy. Figure \ref{30-40_seg_result} shows more mature and accurate segmentations for 30\% and 40\% training data, with some minor boundary deviations, while the internal segmentation remains highly consistent.

To more quantitatively investigate strategy performance, Table \ref{tab:test_results} represent  results of 16 examinations with different data proportion along all strategies. Overall, the Nearest Point strategy consistently achieved competitive or superior results in most metrics, particularly excelling at lower dataset percentages. At 10\%, it achieved the highest Dice (0.9450) and IoU (0.8988) scores, indicating more accurate segmentation performance with limited training data. Similarly, at 20\%, it maintained the best Dice (0.9489), IoU (0.9046), Recall (0.9670), and Hausdorff Distance (HD = 12.5), suggesting improved boundary precision and segmentation stability. For larger dataset portions (30\% and 40\%), performance differences between strategies diminished, showing that model performance converges as more data becomes available. However, at 40\%, the Nearest Point approach again delivered the best Dice (0.9614) and IoU (0.9273), while the Breast Size strategy achieved the lowest HD (16.5), indicating better spatial consistency. 

\begin{table}[ht]
\centering
\caption{Performance metrics of four strategies across different test dataset splits.}
\label{tab:test_results}
\begin{tabular}{c l c c c c c}
\toprule
\textbf{Dataset(\%)} & \textbf{Strategy} & \textbf{Dice}$_{\text{avg}}$ & \textbf{IoU}$_{\text{avg}}$ & \textbf{Precision}$_{\text{avg}}$ & \textbf{Recall}$_{\text{avg}}$ & \textbf{HD}$_{\text{avg}}$  \\
\midrule
\multirow{4}{*}{10} 
    & Random Selection & 0.9404 & 0.8899 & 0.9362 & 0.9474 & 21.6 \\
    & Nearest Point    & \textbf{0.9450} & \textbf{0.8988} & \textbf{0.9547} & 0.9389 & 20.8 \\
    & Breast Size      & 0.9418 & 0.8924 & 0.9338 & \textbf{0.9527} & 22.5 \\
    & Hybrid           & 0.9204 & 0.8585 & 0.9013 & 0.9469 & \textbf{19.8} \\
\midrule
\multirow{4}{*}{20} 
    & Random Selection & 0.9484 & 0.9037 & 0.9481 & 0.9511 & 17.6 \\
    & Nearest Point    & \textbf{0.9489} & \textbf{0.9046} & 0.9340 & \textbf{0.9670} & \textbf{12.5} \\
    & Breast Size      & 0.9474 & 0.9037 & \textbf{0.9587} & 0.9397 & 22.7 \\
    & Hybrid           & 0.9477 & 0.9038 & 0.9558 & 0.9430 & 22.9 \\
\midrule
\multirow{4}{*}{30} 
    & Random Selection & \textbf{0.9563} & \textbf{0.9183} & 0.9548 & 0.9598 & 21.9 \\
    & Nearest Point    & 0.9549 & 0.9156 & 0.9556 & 0.9563 & 23.1 \\
    & Breast Size      & 0.9553 & 0.9161 & 0.9520 & \textbf{0.9607} & 20.7 \\
    & Hybrid           & 0.9506 & 0.9081 & \textbf{0.9606} & 0.9432 & \textbf{20.6} \\
\midrule
\multirow{4}{*}{40} 
    & Random Selection & 0.9497 & 0.9064 & 0.9366 & \textbf{0.9660} & 17.8 \\
    & Nearest Point    & \textbf{0.9614} & \textbf{0.9273} & \textbf{0.9692} & 0.9550 & 23.3 \\
    & Breast Size      & 0.9590 & 0.9227 & 0.9580 & 0.9616 & \textbf{16.5} \\
    & Hybrid           & 0.9537 & 0.9141 & 0.9641 & 0.9459 & 25.1 \\
\bottomrule
\end{tabular}
\end{table}

\begin{figure}[h!]
\centering
\includegraphics[width=1\textwidth, trim=0 0 0 0 , clip]{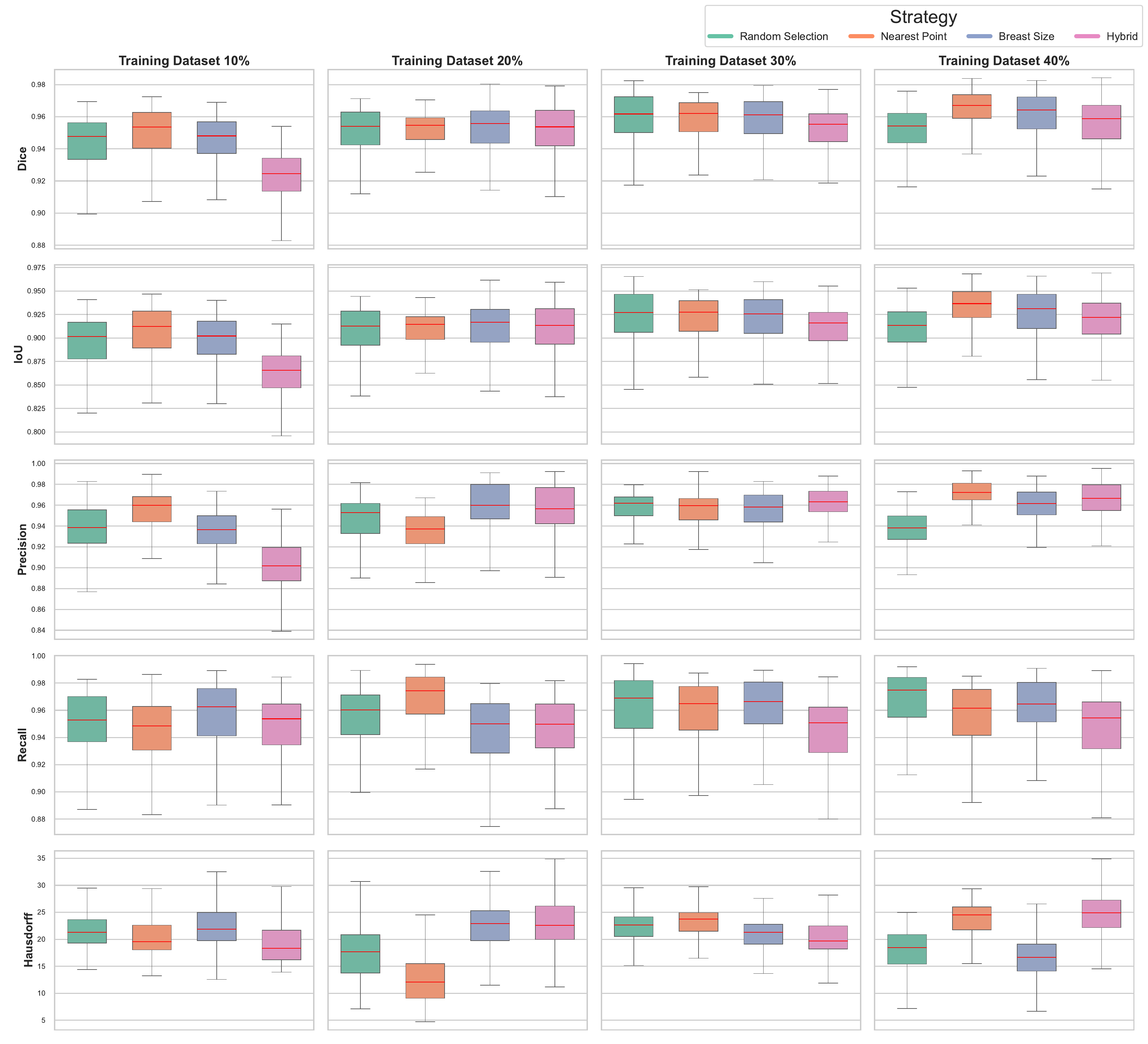}
\caption{\centering Performance metrics distribution across training data percentage and strategies}\label{performance_distribution}
\end{figure}

Since the results in the table represent only the average per slice and do not capture the distribution of the metrics, Figure \ref{performance_distribution} illustrates the performance distribution for each strategy across four training dataset proportions. Specifically, it shows the metric performance for 10\%, 20\%, 30\%, and 40\% training data for the mentioned strategies. As shown inthe figure, both Dice and IoU scores showed a clear upward trend with increasing training set sizes, indicating better overlap between predicted and ground-truth regions. The Nearest Point and Breast Size strategies consistently achieved higher median values than Random Selection and Hybrid, particularly from 20\% onward. Precision exhibited a similar trend, with top-performing strategies showing tighter distributions and reduced variability as the dataset size increased.
In contrast, Recall displayed greater variability at smaller dataset sizes but stabilized with larger training proportions, reflecting improved model sensitivity. The Hausdorff distance decreased as more training data was used, indicating more accurate boundary predictions. 

\subsection{Carbon Emissions and Computational Cost}\label{sec3}
\begin{figure}[h!]
\centering
\includegraphics[width=1\textwidth, trim=0 30 0 40 , clip]{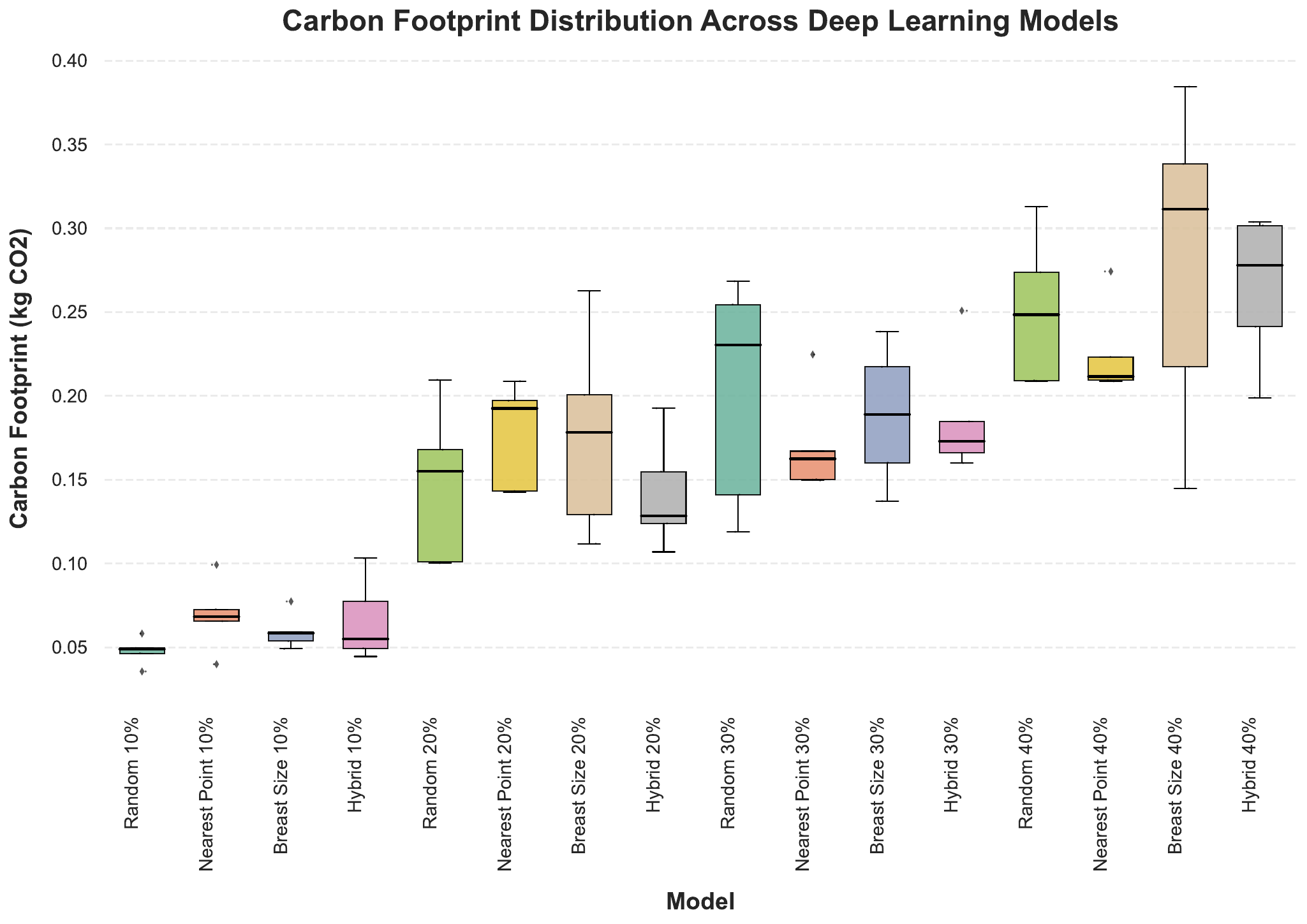}
\caption{\centering Carbon footprint across folds for various strategies}\label{carbon_footprint}
\end{figure}

Carbon Emisions due to AI model training has recently emerged as one of the key factors in DL model sustainabilities \cite{strubell2020energy,zhong2024impact,tamburrini2022ai}. Therefore, it is important to evaluate energy consumption and as a result carbon footprint of AI modelling associated with differnt strategies with different data size. 
Figure \ref{carbon_footprint} illustrates the estimated carbon footprint, expressed in kilograms of CO2, for each data selection strategy across different training dataset sizes (10\%, 20\%, 30\%, and 40\%). As expected, the overall carbon emissions increase with larger training datasets, reflecting the greater computational load and energy consumption required for model optimization on more extensive data. At smaller dataset sizes (10\% and 20\%), all strategies maintain relatively low emissions, though minor differences are already observable. The Hybrid strategy demonstrates the lowest median footprint at these levels, indicating higher training efficiency, while Nearest Point and Breast Size show slightly higher but stable emissions. At 30\% and 40\% training data, variability in emissions becomes more pronounced. Random Selection exhibits the highest and most dispersed carbon footprint values, suggesting inefficient model training and potentially redundant data usage. Conversely, the Nearest Point and Breast Size strategies maintain moderate median footprints with narrower interquartile ranges, implying more stable and computationally efficient training performance. The Hybrid strategy, though efficient at lower dataset sizes, displays a moderate increase in emissions at higher proportions, potentially due to its more complex data selection process.

\section{Discussion}\label{sec4}
The present study provides a comprehensive evaluation of different strategies for the active learning approach. The DL model used in this research is FCN-ResNet50, an eco-friendly architecture for BRS. The findings offer valuable insights into BRS while also highlighting the potential for reducing carbon emissions.

As shown in Figure \ref{histogram_analysis}, the histogram distributions of both Breast Size and Nearest Point distances vary considerably across the dataset. These variations significantly influence the segmentation performance. For instance, in the Random Selection strategy, if the chosen samples fail to adequately represent the full distribution of Breast Size and Nearest Point distances, the model performance can deteriorate compared to other methods. As an example, if only small Breast Sizes located far from the reference coordinate are selected, the random strategy may yield inferior results. This figure clearly demonstrates how variations in Breast Size and placement within MR images affect performance and emphasizes the necessity of ensuring comprehensive distribution coverage in the selected data for active learning. This behavior is also evident in the evolution of the Random Selection strategy shown in Figures \ref{10-20_seg_result} and \ref{30-40_seg_result}. With only 10\% of the data, the random method lacks sufficient breast-position variability, leading to poorly predicted tail regions. As more data are introduced, segmentation quality improves noticeably, particularly for chest wall segmentation and boundary detection.

As presented in the table \ref{tab:test_results}, the Nearest Point strategy demonstrates stronger predictive performance than other strategies across all data percentages, particularly for internal segmentation. However, despite its superior internal performance, it shows weaker boundary detection as reflected by the HD metric. This limitation is primarily attributed to insufficient boundary-focused samples. Nevertheless, this issue is acceptable since it does not significantly affect downstream tasks such as lesion segmentation. Among all strategies, the Random Selection method is the only one that fails to achieve the best precision. This is further supported by the figures, where a high number of false positives (FPs) are observed for this strategy. Although precision varies noticeably among the strategies, recall remains relatively stable, indicating that the large differences in precision stem from FP behavior rather than false negatives (FNs). The Intersection over Union (IoU) follows a similar trend to precision. Overall, all strategies demonstrate high true positive (TP) rates, with only minor differences among them, indicating that the primary challenge lies in balancing FPs and FNs.

The distribution of model performance metrics further substantiates these findings in more detail, as illustrated in Figure \ref{performance_distribution}. A narrower performance distribution across slices reflects higher reliability, which is likely due to fewer FPs and FNs. As the training data proportion increases, the performance distribution becomes increasingly narrow, indicating enhanced stability. This demonstrates the model’s robustness across different slices, from early noisy images where the breast region is not clearly formed to later slices with a fully visible breast structure.

CFP is another critical factor considered in this study. For the 30\% and 40\% data scenarios, the Nearest Point strategy produces the lowest CFP, indicating that the model converges more efficiently under this approach. In contrast, for 10\% and 20\% data sizes, CFP varies significantly but remains substantially lower than training on the full dataset, while still achieving acceptable performance.

Since this research pioneers a combined investigation of extreme CFP reduction and effective segmentation performance, several directions remain open for future work. The BAG analysis only explores two geometric features within MR images that influence model design and data efficiency; therefore, further exploration of advanced feature extraction could enhance active learning performance. Additionally, more sophisticated approaches such as uncertainty-based strategies may further improve results. Another promising avenue for future research is annotation governance, which could significantly influence segmentation accuracy, particularly for chest wall boundary detection. In the current dataset, no standardized guideline exists for defining the precise annotation extent from the chest wall midpoint or for characterizing lesion location uncertainty. Establishing clearer annotation standards could greatly improve consistency and performance. Finally, it is worth to explore alternative DL architectures to determine whether training with very limited data can achieve faster and more stable convergence, as some training instabilities and slow convergence were observed at low data percentages.

\section{Conclusion}\label{sec5}
In this study, an environmentally friendly DL model was employed within an active learning framework for whole-breast segmentation, with a focus on optimizing both performance and sustainability. Since sample selection is the core component of the active learning cycle, choosing an appropriate strategy proved critical for final model performance. Four sampling strategies namely Random Selection, Nearest Point, Breast Size, and a Hybrid approach, were evaluated across four training data proportions (10\%, 20\%, 30\%, and 40\%), with the selected subset used for training and the remaining data for testing to assess generalization. The results demonstrated that increasing the training data proportion generally improved segmentation performance, although Random Selection did not consistently benefit from additional data due to potentially unfavorable sampling distributions. Among all strategies, Nearest Point consistently achieved the highest performance across almost all metrics and training sizes, highlighting its suitability for future breast lesion segmentation applications. Additionally, the Nearest Point strategy exhibited a considerably lower carbon footprint at 30\% and 40\% training data, emphasizing its efficiency and environmental sustainability. Overall, combining the Nearest Point strategy with 30\% of the dataset provided an optimal balance between segmentation accuracy, training efficiency, and minimal carbon emissions, making it a highly effective and sustainable approach for BRS tasks.

\section*{Declarations}
\subsection*{Ethics compliance}
All patients enrolled in the IMAGINE project cohorts received approval from the regional ethics committee. All methods were conducted in accordance with relevant guidelines and regulations, and informed consent was obtained from all participants and their legal guardians.

\subsection*{Data availability}
The Stavanger dataset analyzed in this study contains sensitive patient information and therefore not publicly available. It will be available upon reasonable request by contacting Endre Grøvik through the institution.

\subsection*{Code availability}
The code for data processing, analyzing and modelling is available on GitHub. To access the code repository, please follow the link on \href{https://github.com/SamNarimani-lab/Breast.git}{GitHub.}

\subsection*{Acknowledgment}

\subsection*{Author contributions}
\textbf{Sam Narimani}: Drafted the introduction, methods and materials, proposed strategies for active learning, and authored the results, discussion, and conclusion sections. Additionally, he contributed programming code. \\
\textbf{Solveig Roth Hoff}: Contributed to the writing of the introduction and providing major insights and revision.\\ 
\textbf{Kathinka Dæhli Kurz}: Responsible for data acquisition and preparation, and contributed to the discussion section. \\
\textbf{Kjell-Inge Gjesdal}: Set up MRI protocols. \\
\textbf{Jürgen Geisler}: Support to funding acquisition, review and approval of manuscript \\
\textbf{Endre Grøvik}: Supervised the project, acquired funding and revision draft.




\end{document}